# Genuine driving voltage on polarization fatigue in (Pb,La)(Zr,Ti)O$_3$ antiferroelectric thin films

Wenping Geng,[1] Xiaojie Lou[1*], Jianghong Xu,[1] Fuping Zhang,[2] Yang Liu,[3] Brahim Dkhil,[3] Xiaobing Ren,[1] Ming Zhang,[1] and Hongliang He[2]

[1]*Multi-disciplinary Materials Research Center, Frontier Institute of Science and Technology, Xi'an Jiaotong University, Xi'an 710054, China*

[2]*National Key Laboratory of Shock Wave and Detonation Physics, Institute of Fluid Physics, CAEP, Mianyang 621900, China*

[3]*Laboratoire Structures, Propriétés et Modélisation des Solides, UMR 8580 CNRS-Ecole Centrale Paris, Grande Voie des Vignes, 92295 Châtenay-Malabry Cedex, France*

**Abstract**

The polarization fatigue in (Pb$_{0.97}$La$_{0.02}$)(Zr$_{0.95}$Ti$_{0.05}$)O$_3$ (PLZT) antiferroelectric thin films deposited onto silicon wafers is studied by investigating the effect of the *peak/average/effective* cycling voltage through varying the waveform of the electrical excitation. Interestingly, it is found that the fatigue endurance of the film is determined by the *effective* voltage of the external driving excitation rather than by the *peak* or *average* voltages. Our results can be well explained in the framework of the local phase decomposition model and indicate that the *effective* voltage should be considered as the genuine driving voltage determining the polarization fatigue in PLZT antiferroelectric films.

---

[*] Author to whom correspondence should be addressed; email: xlou03@mail.xjtu.edu.cn or xlou03@gmail.com

Since the discovery of antiferroelectricity in lead zirconate, antiferroelectric materials have been intensively studied for a few decades due to their potential applications in a variety of electrical and microelectronic devices, such as high-energy storage devices, electrical-mechanical transducers, high-strain actuators, and so forth.[1-5] Recently, with the development of microelectronic technologies, thin and thick antiferroelectric films have been also explored for potential microactuator applications in microelectromechanical systems (MEMS).[6,7] For all these applications mentioned above, antiferroelectric devices often operate under a high external ac field for numerous electrical cycles. As a result, high fatigue resistance is required for these devices to ensure their long-term reliability during operation.

Unfortunately, like their ferroelectric counterparts, antiferroelectric materials also display polarization fatigue under high ac electric field, which is a major drawback impeding their commercial and industrial applications. Although a great number of experimental investigations have been devoted to polarization fatigue in ferroelectric materials,[8-10] such work on antiferroelectrics remains scarce. In particular, there are only a few models and scenarios proposed for the origin of polarization fatigue in antiferroelectric materials, all of them being based on those developed for ferroelectrics. The explanations and mechanisms invoked include mechanical degradation via cracks and/or microcracks,[11] domain pinning by agglomerates of charged species,[12,13] reduced stress and degradation due to 180° phase switching,[14] and the local phase decomposition induced by switching-induced charge injection (hereafter referred to as the LPD-SICI model).[15,16] Therefore, there is no consensus on the origin of fatigue in antiferroelectric materials and this phenomenon still remains a matter of debate.

Moreover, it is well known that polarization fatigue in both antiferroelectric and ferroelectric materials generally becomes more severe under higher driving voltages.[8,17-21] In most of such experiments, the effect of the *peak* value (or *magnitude*) of the voltage of the bipolar cycling signal with fixed waveform has been considered for the polarization fatigue studies. However, less attention has been paid to the influence of the waveform (or profile) of the driving signal, though it may also affect the polarization fatigue in a subtle manner.[22,23] Since the driving excitation for the electrical fatigue is characterized not only by its *peak* voltage, but also by its waveform, here we combine both features in order to determine the key parameter, which may be representative of the *genuine* strength of the cycling voltage. As a result, by measuring the fatigue endurance of antiferroelectric $(Pb_{0.97}La_{0.02})(Zr_{0.95}Ti_{0.05})O_3$ (PLZT) films, we show that the *effective* voltage (i.e. the root-mean-square voltage) is the determining parameter and thus may represent the genuine strength of the driving waveform in contrast to both the *peak* and *average* voltages.

A typical antiferroelectric composition of $(Pb_{0.97}La_{0.02})(Zr_{0.95}Ti_{0.05})O_3$ (PLZT) was chosen in the present work. The PLZT thin films of ~400 nm in thickness were deposited on Pt(111)/Ti/SiO$_2$/Si(100) substrates using sol-gel spin coating method.[24] A 0.3 M PLZT precursor solution was prepared from lead acetate trihydrate, lanthanum acetate hydrate, zirconium propoxide and titanium isopropoxide, with 20 mol % Pb-excess in order to compensate Pb losses during thermal treatment. Several PLZT layers were prepared by spin-coating the precursor solution onto substrates at 3000 rpm for 20 s. Each layer was heated at 450 °C for 10 min. A final capping layer of PbO was added and the films were finally annealed at 700 °C for 30 min.[25] The top electrodes (palladium squares of ~100x100

μm$^2$) were deposited through transmission electron microscopy grids by dc sputtering. X-Ray diffraction measurements indicated that the films showed a pure perovskite phase with a highly (100) preferred orientation. The thickness of the PLZT films was measured using cross-section scanning electron microscope (SEM). Polarization fatigue measurements were carried out using an excitation voltage having triangle/sine/square waveform of 100 kHz.[26] After a certain number of fatigue cycles were applied, the hysteresis-loop measurements were conducted using a triangle waveform of 1 kHz. Both the fatigue and hysteresis-loop measurements were carried out using a Radiant technology ferroelectric tester.

From the mathematic point of view, the average value $f_{av}$ and the root-mean-square value $f_{rms}$ can be determined from a periodic function $f(t)$ with the period $T$. Indeed, the average value $f_{av}$ is defined as:

$$f_{av} = \frac{1}{T} \int_0^T f(t)dt \qquad (1)$$

where $t$ is time. We can readily see that $f_{av}(t)$ represents the area underneath the function $f(t)$ divided by $T$. Besides, the root-mean-square value $f_{rms}(t)$ is defined as:

$$f_{rms} = \sqrt{\frac{1}{T} \int_0^T f^2(t)dt} \qquad (2)$$

The schematic diagrams of the driving voltage with triangle, sine, and square waveforms are shown in Fig 1(a)-1(c), respectively. In the present work, the effect of the same *peak* voltage ($V_m$), the same *average* voltage ($V_{av}$) and the same root-mean-square voltage ($V_{rms}$) of the driving signal with triangle/sine/square wave on the fatigue behavior of the PLZT capacitors is systematically investigated. Note that the root-mean-square voltage ($V_{rms}$) is also called *effective voltage* in engineering physics, which is defined by a dc voltage that would produce the same amount of heat as the ac voltage does in a resistor.

Figures 2(a)-2(c) show the fatigue properties of the PLZT thin film through hysteresis-loop measurements after $10^0$, $10^5$, $10^6$, $10^7$, $10^8$, and $10^9$ cycles under the same *peak* voltage ($V_m$=10 V) with triangle [Fig 2(a)], sine [Fig 2(b)], and square [Fig 2(c)] waveform, Figures 2(d)-2(f) display also the fatigue characteristics of the sample but now under the same *average* voltage ($V_{av}$=10 V) with triangle [Fig 2(d)], sine [Fig 2(e)], and square [Fig 2(f)] waveform. Finally, Figures 2(g)-2(i) show the fatigue behaviour obtained by using the same *effective* voltage ($V_{rms}$=10 V) with triangle [Fig 2(g)], sine [Fig 2(h)], and square [Fig 2(i)] waveform. As one can see, both the reduction in the saturated polarization ($P_s$) and the increase in the remanent polarization ($P_r$) demonstrate that polarization fatigue indeed occurs in our PLZT thin-film samples. To better appreciate the changes, the normalized saturated polarization $P_s(N)/P_s(0)$, extracted from the curves in Figure 2, versus the number of fatigue cycles $N$ is plotted in Figure 3, for the same *peak* voltage [Fig 3(a)], the same *average* voltage [Fig 3(b)], and the same *effective* voltage [Fig 3(c)]. In order to ensure the repeatability of our results shown in Figure 2, each measurement was repeated at least three times on virgin capacitors taken on the same film. We found that the difference between the results coming from each measurement is negligible, indicating that the fatigue behaviours shown in Figure 2 are indeed representative of the real features of our PLZT thin film.

From Figures 2(a)-2(c), when the same *peak* voltage ($V_m$=10 V) is used, the loss in $P_s$ after $10^9$ electrical cycles is ~1.6%, ~2.3% and ~12.6% for triangle, sine and square waves, respectively, as depicted in Figure 3(a). Therefore, square waves give rise to more severe fatigue than triangle and sine waves when the *peak* voltage is fixed, which is in good agreement with the results on P(VDF-TrFE) ferroelectric polymer thin films.[23]

Let us now focus on the fatigue behaviour of the sample under the same *average* voltage ($V_{av}$=10 V) with triangle/sine/square waveform [Fig 2(d)-2(f)]. Since the frequency for these three waveforms is fixed (i.e., $10^5$ Hz), the application of the same *average* voltage implies that the areas underneath the triangle/sine/square waveforms for one period are exactly the same. As a consequence, for a fixed *average* voltage (here $V_{av}$=10 V) the *peak* voltages for each waveform are very different, and can be readily calculated using Eq (1) and are reported in Table 1. In our case, $V_m$=20 V, $V_m$=15.7 V and $V_m$=10 V for triangle, sine and square waves, respectively. From Figures 2(d)-2(f) and 3(b), it can be seen that after $10^9$ cycles, the triangle wave function gives rise to the highest fatigue rate with a $P_s$ loss of 45% while square waves lead to the lowest fatigue rate with 12.6% of loss in $P_s$ and sine waves to something in between i.e. 22.3% of loss in $P_s$ (see Table 1). Interestingly after $10^9$ repetitive bipolar cycles, the double-hysteresis-loop characteristic of the virgin antiferroelectric PLZT film becomes seriously suppressed and a rather single hysteresis loop resembling to that of a ferroelectric appears [see figure 2(d)]. Actually, similar results have been also observed in Pb(Nb,Zr,Sn,Ti)$O_3$ antiferroelectric thin films deposited on Pt/Ti/SiO$_2$/Si substrates.[20] We believe that the appearance of a ferroelectric-like hysteresis loop may stem from the following reason: Firstly, the charge injection is induced by phase switching from an antiferroelectric phase to a ferroelectric phase or vice versa under repetitive electrical cycling, and this causes the local phase decomposition at the interface between the film and the electrode. Consequently, the applied field seen by the PLZT layer (free of decomposition) decreases dramatically and becomes unable to switch the sample as the thickness of the interfacial degraded layer grows up with cycling number. It is worth recalling here that in any antiferroelectric, above the coercive field (i.e. when the loops occur), the applied electric voltage will induce a phase transition from an antiferroelectric to a ferroelectric phase, indicating that both phases are very close in the energy landscape. Secondly, the stress/strain and/or the new electrical boundary conditions induced by local phase decomposition during fatigue may also stabilize the ferroelectric phase at a zero external field. Other explanations proposed in the literature[20,27] for this phenomenon invoke also domain-wall pinning or accumulation of oxygen vacancies on Pt electrodes under high driving voltage.

Let us come back to both figure 2 and figure 3. It is clear that by fixing the *peak* or *average* voltage value, the fatigue endurance of the PLZT sample varies considerably when the waveform considered is changed (Table 1). Therefore, neither the *peak* voltage nor the *average* voltage is adapted for determining the fatigue properties. In contrast [Fig 2(g)-2(i) and Fig 3(c)], the *effective* voltage, which is actually a combination of both the amplitude and the waveform of the external stimulus in a specific form [see Eq (2)], gives rise to very similar fatigue rates, i.e., ~12.5% of loss in $P_s$ after $10^9$ bipolar cycles, by taking the same value of $V_{rms}$=10 V and that whatever the waveform considered (see Table 1). Therefore, the *effective* voltage by including both features i.e. amplitude and shape of the applied cycling voltage, appears to be the best parameter to be used in polarization fatigue studies and may thus represent the genuine strength of the driving voltage.

For better illustration of the effect of the *peak/average/effective* voltage on polarization fatigue in our PLZT samples, all the driving voltage used in Fig 2(a)-2(i) are converted into their *peak/average/effective* values in Table 1, according to Eqs (1) and (2). The normalized reduction in the saturated polarization, denoted by $\triangle P_s/P_s(0)$ where $\triangle P_s=$ $P_s(0)$- $P_s(10^9)$, is also shown for comparison. From Table 1, one can clearly see that there is NO correlation between the *peak* or the *average* voltage value and $\triangle P_s/P_s(0)$. In contrast, a strong relationship appears between the *effective* voltage and $\triangle P_s/P_s(0)$, that is, the higher the *effective* voltage, the more severe the losses in the saturated polarization will be.

Our findings in the present work raise an interesting and important question: why is it the *effective* voltage that governs the fatigue behavior? As mentioned above, the *effective* voltage of an ac waveform is defined as a dc voltage that could generate the same amount of heat in a resistor as the ac waveform itself does. Therefore, the *effective* voltage is closely related to thermal power or heat generated during repetitive electrical cycling during fatigue measurements. Referring to the fatigue models and scenarios proposed in the literature, no one explicitly invokes the relationship between polarization fatigue and heat generation, except the LPD-SICI model.[8,10,15,16,28] So, let us now give a tentative interpretation of the phenomena observed in this work using the LPD-SICI theory.

From the viewpoint of the LPD-SICI model, polarization fatigue is caused by local phase decomposition arising from switching-induced charge injection at the electrode-sample interface.[8,10,28] In this model, there is a critical parameter $E_{bc}J$, called the local injected power density or local heat generated by injected electrons (see Ref [10] and Ref [15] for more details), in which $E_{bc}$ is the depolarization field caused by the head-to-head or tail-to-tail bound charges at the tip of the needle-like domain (or phase) during domain (or phase) switching under electric field in ferroelectric (or antiferroelectric) materials. For simplicity, assuming the local injected current density $J = \sigma_i^{eff} \cdot E_{bc}$, where $\sigma_i^{eff}$ is the effective conductivity at the electrode-film interface, we obtain $E_{bc}J = \sigma_i^{eff} E_{bc}^2$. Because the film is driven by an ac field $E_{appl}$ with the period $T$ under repetitive electrical cycling, $E_{bc}$ should also be a $T$-periodic function $E_{bc}(t)$. Recalling Eq (2), we can define an effective value (or a root-mean-square value) $E_{bc}^{rms}$ as $E_{bc}^{rms} = \sqrt{\frac{1}{T}\int_0^T E_{bc}^2(t)dt}$. Moreover, it is reasonable to argue that there is a strong correlation between $E_{bc}$ and the external field $E_{appl}$ (or equivalently the applied external voltage $V$): the higher $E_{appl}$ (or $V$), the larger $E_{bc}$ will be. Therefore, a higher *effective* voltage $V_{rms}$ implies a higher effective field $E_{bc}^{rms}$ at the domain/phase nucleation site[10,15] and thus should guarantee a large amount of heat generated locally, and consequently a higher fatigue rate. Therefore, the experimental results shown in this work can be well explained in the framework of the LPD-SICI model and thus indicate that polarization fatigue in antiferroelectric materials and as a matter of fact in ferroelectrics has to do with heat generation during repetitive electrical cycling voltages. It is worth mentioning that the local phase decomposition (or even partially melting) during fatigue process in ferroelectric materials has been *in situ* experimentally observed using micro-Raman spectroscopy[28] or using SEM.[29] These observations in addition to our experimental evidences provide thus a strong support for considering much more the LPD-SICI model in fatigues properties.

Finally, let us add a few remarks on the data shown in Fig 2(a), Fig 2(b) and Fig 3(a). From these figures using the same *peak* voltage of 10 V, triangle and sine waves give rise to a negligible fatigue rate in comparison with square waves. Referring to the double hysteresis loop recorded from the virgin capacitor shown in Fig 2, one can see that the antiferroelectric-ferroelectric phase transition voltage $V_{AFE-FE}$ is about 7 volt (i.e., ~175 kV/cm). Although the *peak* voltage ($V_m$=10V) of the triangle [Fig 2(a)] and sine [Fig 2(b)] wave is larger than $V_{AFE-FE}$, the *effective* voltage $V_{rms}$ is only 5.8 V (for triangle wave) and 7.1

V (for sine wave), values which are below or close to $V_{AFE-FE}$ of ~7 V, and are the lowest among all the data (see Table 1). Therefore, the low fatigue rates measured for triangle and sine waveforms with $V_m$=10V in Fig 2(a) and 2(b) can be safely attributed to the incomplete switching of the antiferroelectric phase and therefore less heat generated during repetitive cycling. Interestingly, a similar behavior has been observed for ferroelectric capacitors, where the concepts of sufficient (or complete) switching and insufficient (or incomplete switching) were developed.[30]

As a conclusion, by using a systematic and careful analysis of the fatigue endurance of PLZT antiferroelectric films by varying the waveform of the electrical excitation, we demonstrate that neither the *peak* nor the *average* of the cycling voltage is well-adapted to properly study the fatigue properties. In contrast, we show that the *effective* voltage which takes into account all the features of the electrical excitation (amplitude plus shape) is the most appropriate parameter and thus it is revealed to be the genuine driving voltage on the polarization fatigue. We also explained our observations using the LPD-SICI model which highlights that local phase decomposition represents a major mechanism involved in fatigue properties of (anti-)ferroelectric materials and thus deserves further considerations.

This work was supported by National Key Laboratory of Shock Wave and Detonation Physics through a fund (No. LSD201201003) and the Ministry of Science and Technology of China through a 973-Project (No. 2012CB619401). X.J. Lou would like to thank the "One Thousand Youth Talents" program for support.

Figure captions:

Fig 1 Schematic of the driving signal with (a) triangle, (b) sine, and (c) square waveforms, which were used for fatiguing the sample

Fig 2 Polarization fatigue properties of the PLZT antiferroelectric thin films through hysteresis-loop measurements after $10^0$, $10^5$, $10^6$, $10^7$, $10^8$, and $10^9$ cycles under the same *peak* voltage ($V_m$=10 V) with triangle (a), sine (b), and square (c) waveform; (d)-(f) panels show the fatigue behaviour of the sample using the same *average* voltage ($V_{av}$=10 V) with triangle (d), sine (e), and square (f) waveform. And (g)-(i) panels show the fatigue character of the sample under the same *effective* voltage ($V_{rms}$=10 V) with triangle (g), sine (h), and square (i) waveform.

Fig 3 Comparison of $P_s(N)/P_s(0)$ as a function of the number of cycles ($N$) for (a) the same *peak* voltage ($V_m$=10 V), (b) the same *average* voltage ($V_{av}$=10 V), and (c) the same *effective* voltage ($V_{rms}$=10 V) with triangle/sine/square waveform.

Table 1. The *effective* voltages $V_{rms}$ and the corresponding *peak* voltages $V_m$ and *average* voltages $V_{av}$ for triangle/sine/square waveforms, and their relationship with the losses in the saturated polarization $\Delta P_s/P_s(0)$.

| Curve | Waveform | $V_m$(V) | $V_{av}$(V) | $V_{rms}$(V) | $\Delta P_s/P_s(0)$ |
|---|---|---|---|---|---|
| Figure 2 (a) | Triangle | **10** | 5 | 5.8 | 1.6% |
| Figure 2 (b) | Sine | **10** | 6.37 | 7.1 | 2.3% |
| Figure 2 (c) | Square | **10** | 10 | 10 | 12.6% |
| Figure 2 (d) | Triangle | 20 | **10** | 11.5 | 45% |
| Figure 2 (e) | Sine | 15.7 | **10** | 11.1 | 22.3% |
| Figure 2 (f) | Square | 10 | **10** | 10 | 12.6% |
| Figure 2 (g) | Triangle | 17.3 | 8.65 | **10** | 13.9% |
| Figure 2 (h) | Sine | 14.1 | 8.98 | **10** | 11.8% |
| Figure 2 (i) | Square | 10 | 10 | **10** | 12.6% |

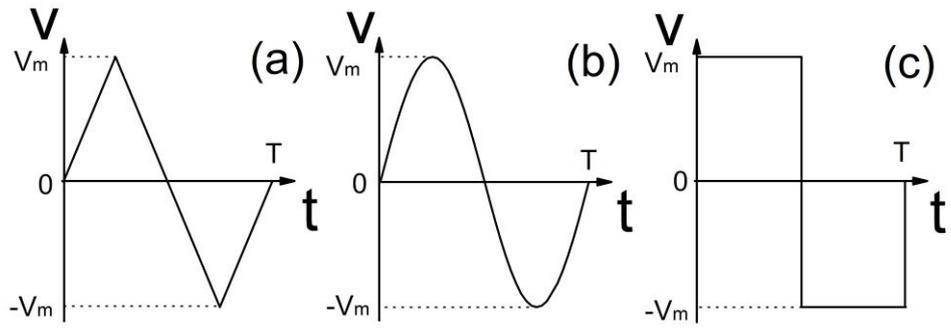

Figure 1

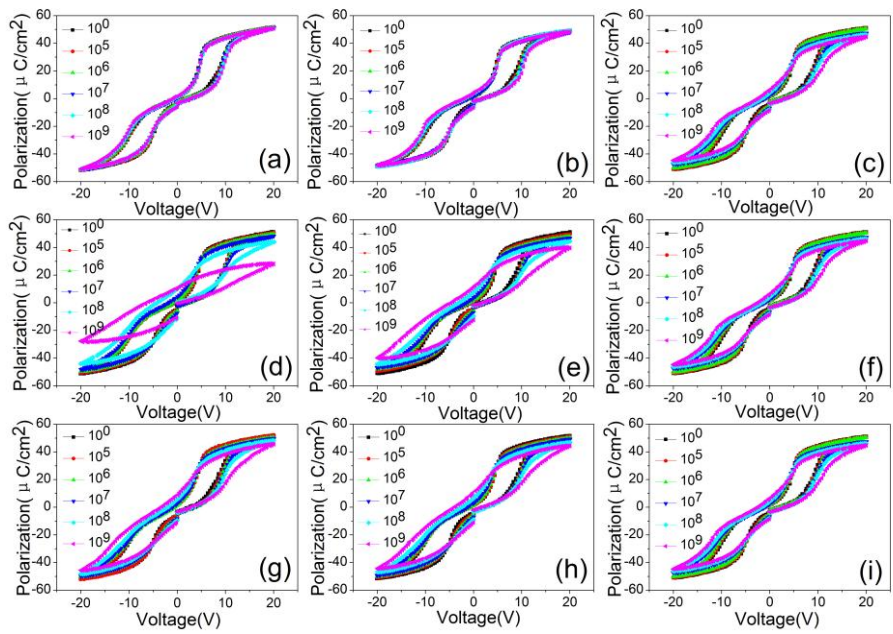

Figure 2

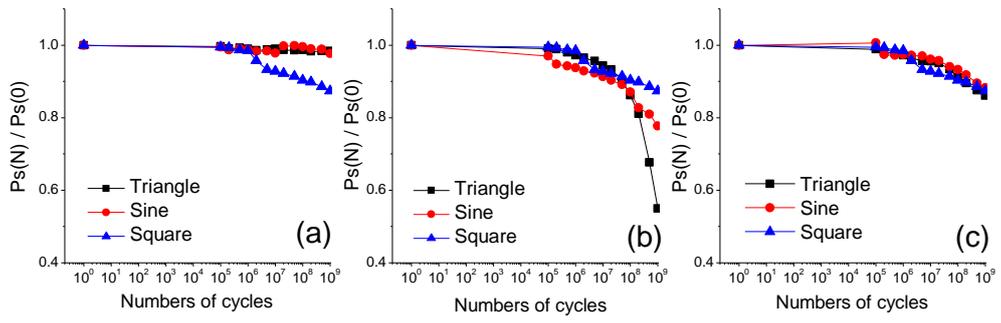

Figure 3